\documentclass{aastex6}

\usepackage{color} 
\graphicspath{{./figures/}} 
\usepackage{color} 

\fullcollaborationName{}

\begin{document}


\title{Dynamical Evolution of the Debris after Catastrophic Collision around Saturn}


\author{Ryuki Hyodo\altaffilmark{1,2}, S{\'e}bastien Charnoz\altaffilmark{2}}


\altaffiltext{1}{Earth-Life Science Institute/Tokyo Institute of Technology, 2-12-1 Tokyo, Japan}
\altaffiltext{2}{Institut de Physique du Globe, Paris 75005, France}

\begin{abstract}
The hypothesis of a recent origin of Saturn’s rings and its mid-sized moons is actively debated. It was suggested that a proto-Rhea and a proto-Dione might have collided recently, giving birth to the modern system of mid-sized moons. It is also suggested that the rapid viscous spreading of the debris may have implanted mass inside Saturn’s Roche limit, giving birth to the modern Saturn’s ring system. However, this scenario has been only investigated in very simplified way for the moment. This paper investigates it in detail to assess its plausibility by using $N$-body simulations and analytical arguments. When the debris disk is dominated by its largest remnant, $N$-body simulations show that the system quickly re-accrete into a single satellite without significant spreading. On the other hand, if the disk is composed of small particles, analytical arguments suggest that the disk experiences dynamical evolutions in three steps. The disk starts significantly excited after the impact and collisional damping dominates over the viscous spreading. After the system flattens, the system can become gravitationally unstable when particles are smaller than $\sim$ 100 m. However, the particles grow faster than spreading. Then, the system becomes gravitationally stable again and accretion continues at a slower pace, but spreading is inhibited. Therefore, the debris is expected to re-accrete into several large bodies. In conclusion, our results show that such a scenario may not form the today's ring system. In contrast, our results suggest that today's mid-sized moons are likely re-accreted from such a catastrophic event.\\

\end{abstract}




\section{Introduction} \label{sec:intro}
Origin, age and dynamical evolution of  icy Saturn's rings and satellites are still debated. \cite{Can10} has proposed that Saturn's rings formed by tidal disruption of a Titan-sized body that migrates inward through the interaction with 
circumplanetary gas disk about $4.5$ Gyrs ago. On the other hand, \cite{Hyo17} showed that tidal disruption of a passing Pluto-sized Kuiper belt object can form ancient massive rings around, not only, Saturn but also other giant planets 
during the Late heavy bombardment (LHB) about $3.8$ Gyrs ago. Then, the inner regular satellite systems around Saturn, Uranus and Neptune are, generally, thought to be formed by spreading of such ancient massive rings \citep{Cha10,Cri12,Hyo15a,Hyo15b}.\\

The pure icy rings would continuously darken over the age of solar system due to micrometeorid bombardment \citep[e.g.][]{Cuz98}. So, the rings might be formed more recently than it has been thought. Note that, however, they might be older if they are more massive \citep{Ell11,Esp12}. 
Recently, \cite{Cuk16} has investigated the past orbital evolutions of Saturn's midsized moons (Tethys, Dione and Rhea) and found that Tethys-Dione 3:2 orbital resonance is not likely to have occurred whereas the Dione-Rhea 
5:3 resonance may have occurred. Then, they conclude that the midsized moons are not primordial and propose that the moons re-accreted from debris disk that formed by a catastrophic collision between primordial Rhea-sized moons 
about 100 Myrs ago \citep{Cuk16}. They also propose that the debris disk may spread inward rapidly (due to fast gravitational instability) and feed the Roche limit to form the today's rings. 
In addition they propose that outward spreading may form and push outward a population of small moons (with a mass of $m=4\times10^{20}$ kg) that would excite Titan's current eccentricity through the resonant interaction.\\

The aim of the present paper is to test this scenario by using direct simulations and detailed analytical arguments. In Section \ref{SPH}, we first use smoothed-particle hydrodynamics (SPH) simulations to investigate the outcome 
of the collision between two proto-Rhea sized objects at impact velocity $3$ km s$^{-1}$ \citep{Cuk16}. In section \ref{N-body}, using $N$-body simulations, we investigate the long-term evolution of the debris, starting from the impact simulation and assuming 
that debris is not collisionally disrupted. In section \ref{analysis}, using analytical arguments, we estimate the fate of disk of small particles as an extreme case of collisional evolution. In section \ref{Conclusion}, we discuss the plausibility of this scenario to form today's rings and moons.\\

\section{Catastrophic collision between Rhea-sized bodies}
\label{SPH}
\subsection{SPH methods and models}
Using SPH simulations, we model collision between Rhea-sized objects ($M_{\rm body}=10^{21}$ kg) in free space. The silicate mass fraction of Saturn's icy moons are diverse \citep{Cha11}. 
Thus, we assume $60$wt\% silicate core for one object and $40$wt\% silicate core for the other with both covered by icy mantel. Following \cite{Cuk16} arguments, impact velocity is set to be about $6$ times of the mutual escape 
velocity which is about $v_{\rm imp}=3$ km s$^{-1}$. Impact angle is set to be either $\theta=0, 20, 45, 60,$ and $80$ degrees. The total mass of the two colliding objects is $M_{\rm tot}=2\times10^{21}$ kg and the total number of SPH particles is $N=2\times10^5$. 
We simulated about $3.88$ hours which is much shorter than the orbital period at the distance of Rhea (4.5 days).  Our numerical code is the same as that used in \cite{Hyo16, Hyo17}, which was developed in \cite{Gen12}.

\begin{figure}[h!]
\figurenum{1}
	\epsscale{0.5}
	\plotone{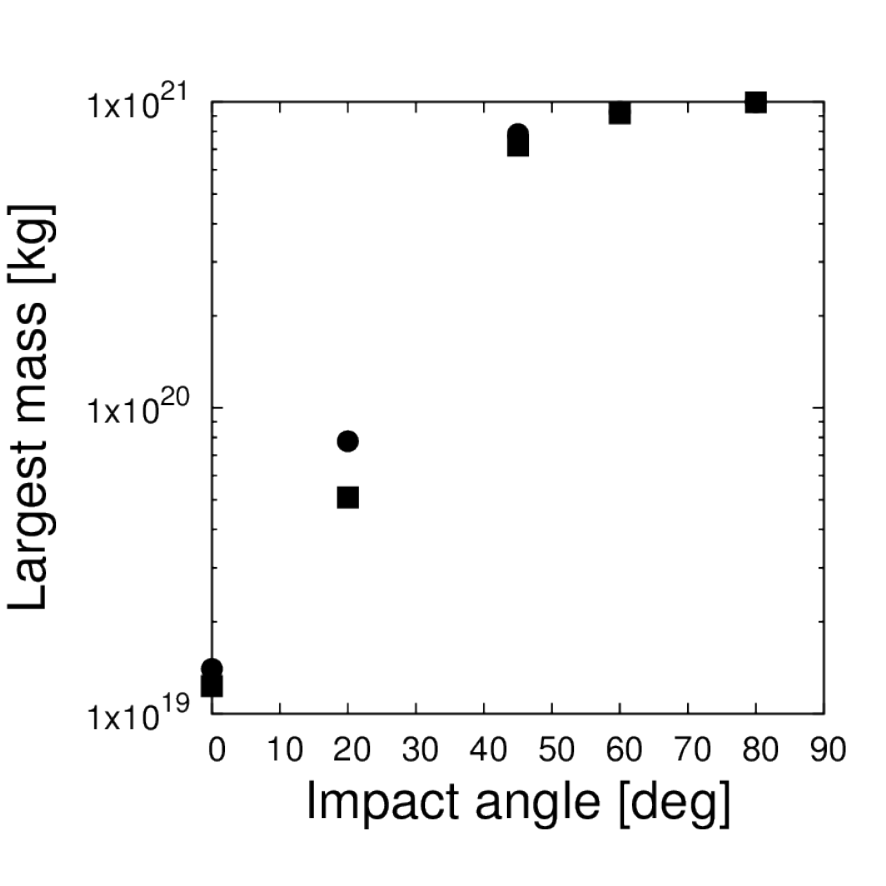}
 \vspace{0mm}
\caption{Largest two remnants after the collision as a function of different impact angles obtained from SPH simulations. Filled circles and squares represent the largest and the second largest fragments, respectively.}
\label{largest_rem}
\end{figure}

\subsection{Results of SPH simulations}
\begin{figure}[ht!]
\figurenum{2}
		\epsscale{0.8}
	\plotone{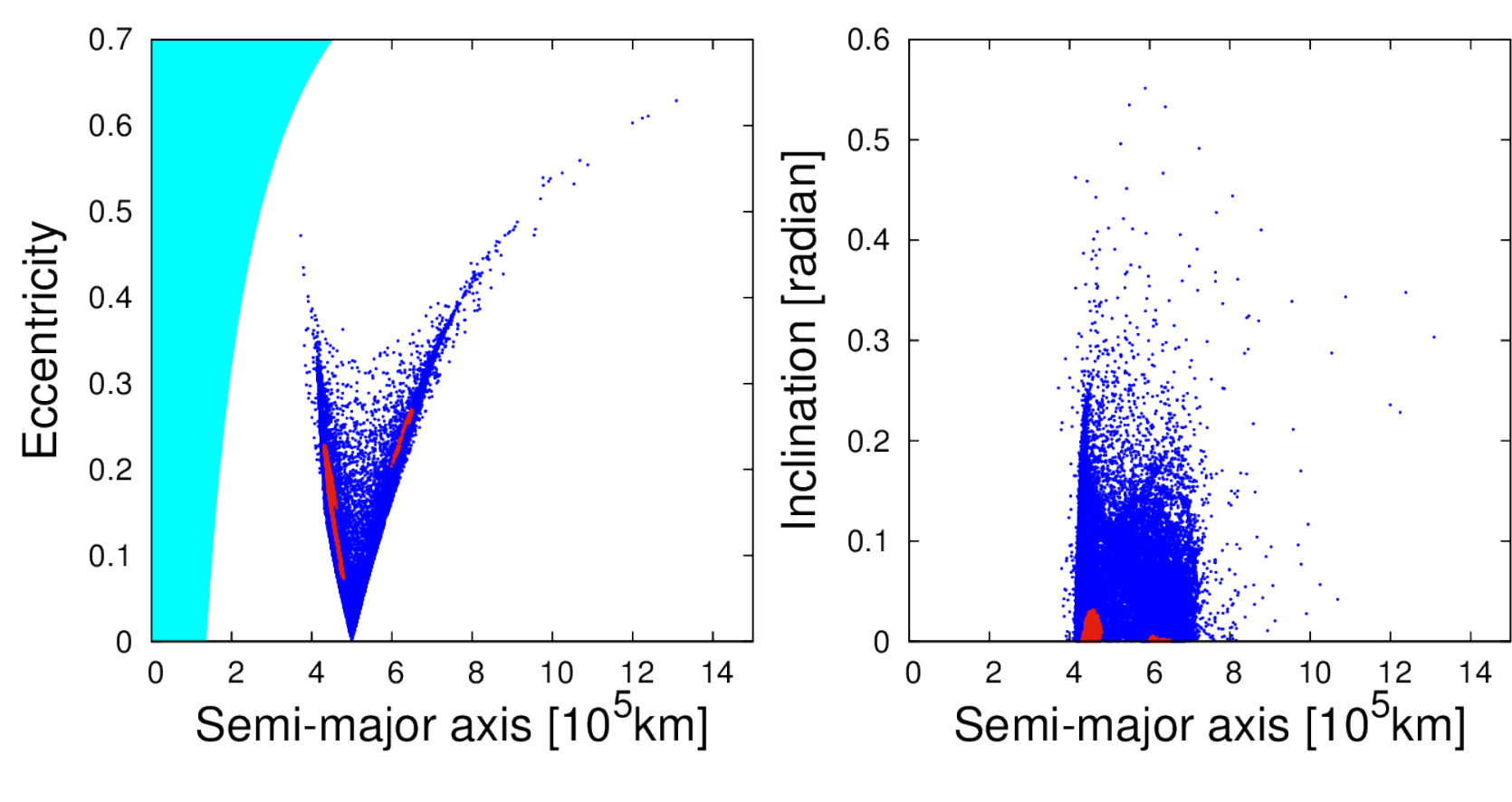}
 \vspace{0mm}
\caption{Orbital elements of the debris after the impact  in the case of $\theta=45$ degrees obtained from SPH simulations, assuming collision takes place at $a=5\times10^5$ km.  
		The left and right panels show eccentricities and inclinations of particles against their semi-major axes. 
		The blue dots represent water ice particles and red dots represent silicate particles. 
		Cyan area on the left panel corresponds to the region inside the Roche limit where $a(1-e)<135,000$ km.}
\label{initial_dis}
\end{figure}

SPH simulations show that the collision is energetic enough to catastrophically destroy colliding objects (Figure \ref{largest_rem}) as suggested by \citep{Cuk16}. 
However, after the collision, in most of cases, two large fragments remain as direct leftovers of the cores covered by water ice of the original two colliding objects. 
In the case of $\theta=45$ degrees, the largest remnants consist of masses of $M=7.8\times10^{20}$ kg  and $M=7.2\times10^{20}$ kg which are both about 40\% of the total mass of the two objects. 
Figure \ref{initial_dis} shows the orbital elements of the debris after the impact in the case of $\theta=45$ degrees, assuming the impact occurs at semi-major axis $a=5 \times 10^5$ km (as in \cite{Cuk16} and used as initial condition for $N$-body simulations (Section \ref{N-body})). 
Initial dispersion of the semi-major axes and eccentricities are about $3.5\times10^5$ km and $0.35$, respectively, which are consistent with what we can derive from the first-order approximation as 
\begin{eqnarray}
	&\Delta a_{\rm ini} \sim 2 \Delta v/\Omega\\
	&\Delta e_{\rm ini} \sim \Delta v/\left(a \Omega \right)
\end{eqnarray}
where  $ \Delta v \sim v_{\rm imp} $ and $\Omega$ are the velocity dispersion and orbital frequency, respectively. In the next section, we investigate the longer-term evolution of the debris.

\section{Dynamical Evolution of Debris with large fragments}
\label{N-body}
\subsection{$N$-body methods and models}
Orbits of the debris are integrated by using a forth-order Hermite method \citep{Mak92}. The collisions between particles are solved as hard-sphere model with the normal and tangential coefficient of restitutions $\epsilon_{\rm n}=0.1$ and $\epsilon_{\rm t}=1$, respectively.  
However, following the argument of \cite{Kok00, Can95}, we allow accretion only when the following two conditions are satisfied. First, the Jacobi energy of two particles after the collision $E_{J}$ has to be negative as
\begin{equation}
	E_{\rm J} = \frac{1}{2}v^2\epsilon_{\rm eff}^2 - \frac{3}{2}x^2\Omega^2 + \frac{1}{2}z^2\Omega^2 - \frac{G\left(m_1 + m_2 \right)}{r} + \frac{9}{2}r_{\rm H}^2\Omega^2  < 0
\end{equation}
where $x, y$ and $z$ are the relative positions with $r^2=x^2+y^2+z^2$, $m_{1,2}$ are the masses of  particles, and $\epsilon_{\rm eff}$ is an effective coefficient of restitution written as
\begin{equation}
	\epsilon_{\rm eff} = \left[ \left( \epsilon^2_{\rm n}v^2_{\rm n} + \epsilon^2_{\rm t}v^2_{\rm t} \right)/\left( v^2_{\rm n} + v^2_{\rm t} \right) \right]^{1/2}
\end{equation}
where $v_{\rm n}$ and $v_{\rm t}$ are the normal and tangential components of the relative velocity between particles. In addition, the sum of the radii of two particles should be smaller than the Hill radius as 
\begin{equation}
	r_1 + r_2 = 3^{1/3}\frac{1+\mu^{1/3}}{\left( 1+\mu \right)^{1/3}} \left( \frac{\rho_{\rm par}}{\rho_{\rm pla}} \right)^{-1/3} \frac{R_{\rm pla}}{a} r_{\rm H} \leq r_{\rm H}
\end{equation}
where $\rho_{\rm pla}$ and $\rho_{\rm par}$ are the densities of planet and particles, respectively. $\mu$ is the mass ration $m_2/m_1$. $R_{\rm pla}$ is the radius of the planet and $r_{\rm H}$ is the Hill radius defined as 
\begin{equation}
	r_{\rm H} = \left( \frac{m_1+m_2}{3M_{\rm pla}} \right)^{1/3}a.
\end{equation}

We use tree-method for the gravity calculations and collisional detections \citep{Rei12,Hyo15a}. The numerical code is the same as that used in \cite{Hyo15a}.\\

\subsection{Initial conditions}
Positions and velocities of particles obtained from SPH simulation with $\theta=45$ degrees (see Figure \ref{initial_dis}) are passed to $N$-body simulations, assuming the collision takes place in the equatorial plane of Saturn and the center of mass of the two colliding objects 
orbits around Saturn with semi-major axis $a=5.0 \times 10^5$ km and the eccentricity $e=0$. We also include Titan with the current semi-major axis  $a_{\rm Titan}=1.2\times10^{6}$ km, eccentricity $e_{\rm Titan}=0.0288$ and inclination $i_{\rm Titan}=0.34$ degrees. 
Due to the computational power limitation, we randomly select $20,000$ particles from 200,000 particles used in SPH simulations. 
We run 5 different simulations by changing the random choise of particles. 
Initially each particle has same mass of $m=M_{\rm tot}/N$ ($m=1\times10^{17}$ kg) and they are either silicate or icy particles. 
We assume silicate particles have density $\rho_{\rm sil} = 3000$ kg m$^{-3}$ and icy particles have $\rho_{\rm icy} = 900$ kg m$^{-3}$.  During the calculation, we track the density change when two particles merge into a new particle. 
Just after the calculations start, numerous particles merge into single particles as they are initially the constituent particles of large remnants. 

\subsection{Results of $N$-body simulaitons}
\begin{figure}[ht!]
\figurenum{3}
	\epsscale{0.56}	
	\plotone{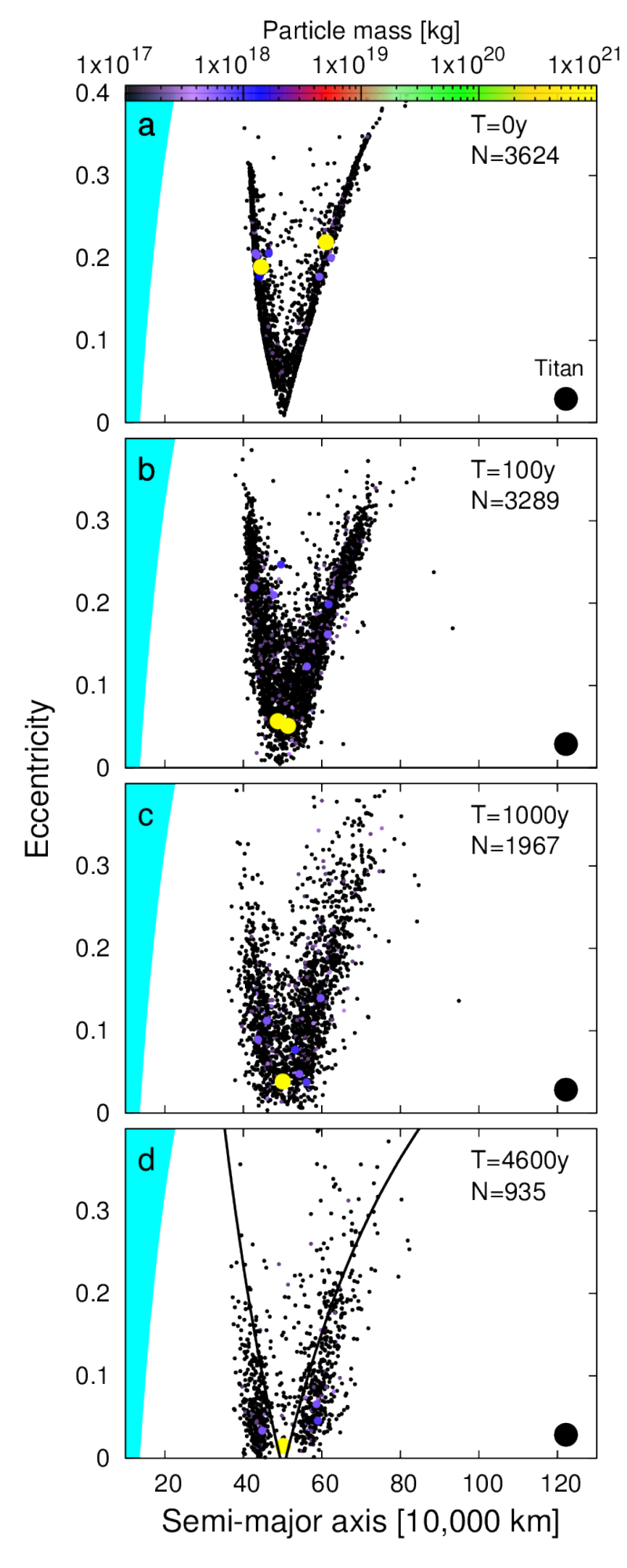}
 \vspace{0mm}
\caption{Time evolution of a debris disk on the $a-e$ plane. The dots represent particles and their color and size represent their mass. The black filled big dot on the bottom right represents Titan. 
	Cyan area on the left panel corresponds to the region inside the Roche limit where $a(1-e)<135,000$ km. 
	Two black lines in panel (d) represent the orbital elements that cross the orbits of the largest remnant at either pericenter or apocenter.}
\label{Nbody_snap}
\end{figure}

Figure \ref{Nbody_snap} shows the time evolution of the system. 
Just after the impact, most of the mass is contained in the two largest remnants (Figure \ref{Nbody_snap}, panel (a)). 
Since the two remnants have large eccentricities ($e\sim0.2$), their orbits cross. 
Thus, after several periods, they collide and merge into a single large body with a mass of $m \sim 1.5 \times 10^{21}$ kg with small eccentricity (Figure \ref{Nbody_snap}, panels (b) and (c)).\\

\begin{figure}[h!]
\figurenum{4}
	\epsscale{0.9}	
	\plotone{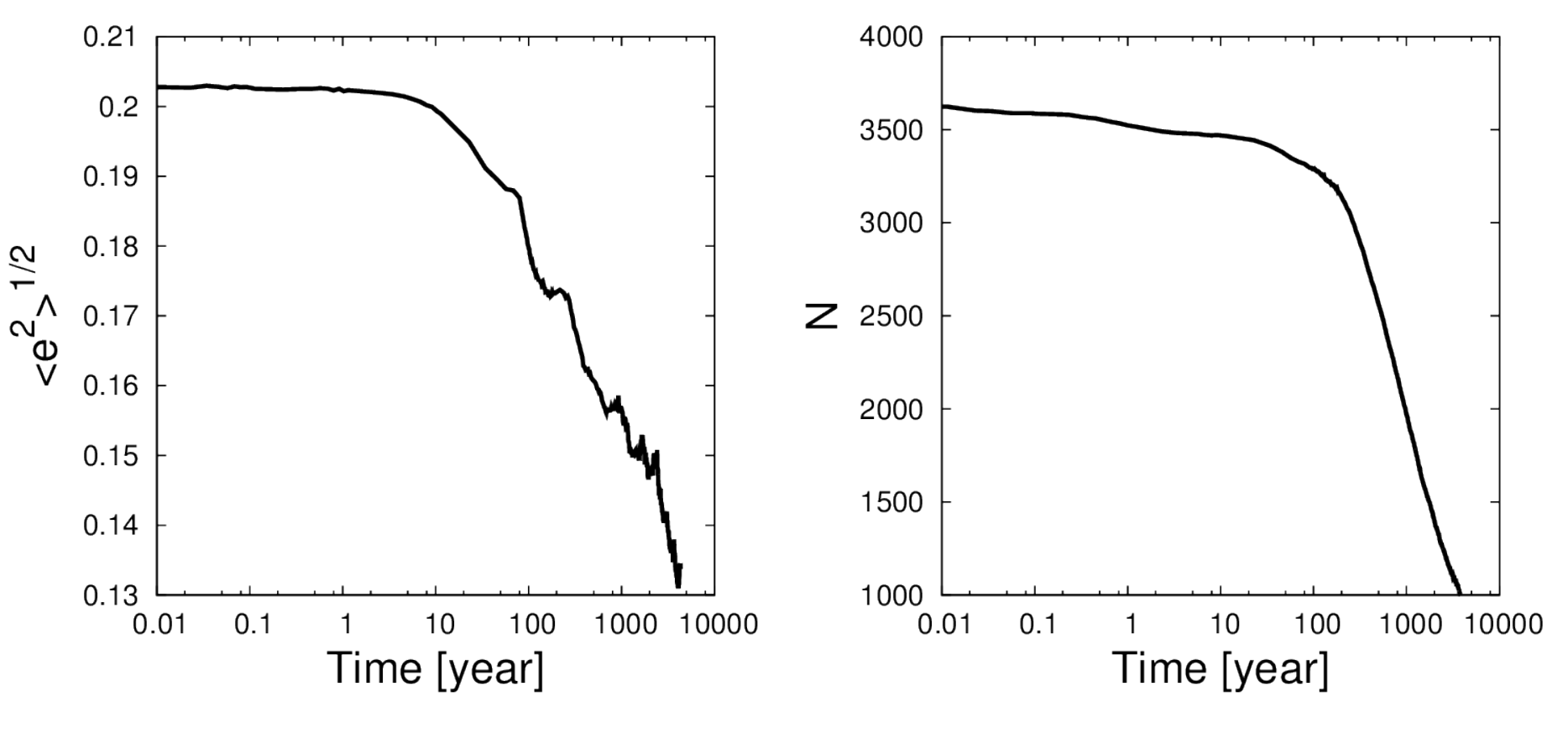}
 \vspace{0mm}
\caption{Evolution of r.m.s eccentricity (left panel) and number of particles in the system (right panel) in the $N$-body simulation after the impact.}
\label{rms_N_evo}
\end{figure}

Mass of most field particles are $m_{\rm p}=10^{17}$ kg and their escape velocity is $v_{\rm esc} \sim 20$ m s$^{-1}$. In order for accretion between such particles to take place, relative velocities should be smaller than 
their escape velocity. Thus, in order to accrete, eccentricity of field particles should be smaller than $e_{\rm cri} \sim 2.5 \times 10^{-3}$. 
Left panel of Figure \ref{rms_N_evo} shows the time evolution of the root mean square (RMS) of eccentricities $\langle e^2\rangle^{1/2}$. 
Since the field particles have much larger eccentricities than $e_{\rm cri}$, accretion between field particles is initially difficult. 
However, collisional damping is effective and the RMS eccentricity decreases with time (Figure \ref{rms_N_evo}, left panel).  
As the largest remnant is much larger than the field particles, field particles whose eccentricities are below $e \sim 0.06$ can accrete onto the largest remnant rather than between themselves.  
We confirm by $N$-body simulations that the remnant keeps growing by eating field particles and the number of particles in the system keeps decreasing (Figure \ref{rms_N_evo}, right panel).\\

At the end of our $N$-body simulations, we have less than $1000$ particles without significant spreading of the system (Figure \ref{Nbody_snap}, panel (d)). At this time, the largest remnant (satellite) has accreted most of the 
field particles whose orbits cross that of the satellite and it has a mass of $m\sim1.9\times10^{21}$ kg ($\sim 95$ \% of the total system mass) with small eccentricity $e \sim10^{-2}$. 
The size of the Hill sphere of this largest remnant is about $5000$ km. The typical separation between two bodies is 10 Hill radius \citep{Kok98}. 
Thus, the remaining field particles would accrete onto the largest remnant and the system is expected to re-accrete into a single large object.  

\newpage
\newpage
\section{Dynamical evolution of debris of small particles}
\label{analysis}
In the previous section, using $N$-body simulations, we investigated the long-term evolution of the debris disk within 
which initially two large fragments are embedded as a result of catastrophic collision. 
However, we neglected the effect of fragmentation and the debris particles initially have large eccentricities, thus collisional grinding may occur in the real system.  
Here, we analytically estimate the fate of the debris, initially consisting of same-sized small particles (radius $r_{\rm p}$). The velocity dispersion $c$ of the system is controlled by the following equation. 
\begin{equation}
	\frac{dc^2}{dt} = f_{\rm trans}c^2 + f_{\rm col} \left(r_{\rm p} \Omega \right)^2 + f_{\rm vs}\frac{v_{\rm esc}^4}{c^2} - f_{\rm damp}c^2
\end{equation}
where the first two terms are the contribution of viscous heating: the first term is due to velocity shear sampled by random motion of particles \citep{GT78} and the second term is due to physical collisions \citep{Ara86}. 
The third term is due to gravitational scattering described by Chandrasekhar's relaxation time \citep{Ida90,Mic16} and the last term is due to collisional damping \citep{GT78}. 
The coefficients are written as 
\begin{eqnarray}
f_{\rm trans}=&c_1 \times \frac{9}{4} \frac{\tau}{1+\tau^2} \Omega\\
f_{\rm col}=&\frac{9}{4}\tau \Omega\\
f_{\rm vs}=&\frac{\Omega \tau \ln \Lambda}{4}\\
f_{\rm damp}=&c_2 \times \Omega \tau \left( 1 -  \epsilon^2 \right)
\end{eqnarray}
where $\tau$ is the optical depth and is written with the assumption that all particles have the same radius $r_{\rm p}$ as 
\begin{equation}
	\tau = \frac{N_{\rm tot}\pi r_{\rm p}^2}{S} 
	\sim 1.1 \times
	\left( \frac{\rho_{\rm p}}{1200 \hspace{0.1em} {\rm kg \hspace{0.1em} m^{-3}}} \right)^{-1} 
	\left( \frac{S}{1.1\times10^{18} \hspace{0.1em} {\rm m^2}} \right)^{-1} 
	\left( \frac{M_{\rm tot}}{2 \times10^{21} \hspace{0.1em} {\rm kg}} \right) 
	\left( \frac{r_{\rm p}}{1.0 \hspace{0.1em} {\rm m}} \right)^{-1} 
\end{equation}
where $N_{\rm tot}$ is the total number of particles, $\rho_{\rm p}$ is the particle density and $S$ is the surface area, respectively. 
Assuming $\rho_{\rm p}=1200$ kg m$^{-3}$ and $S=2\pi a \Delta a = 2\pi \times (5\times10^8 {\rm m}) \times (3.5\times10^8 {\rm m}) \sim 1.1\times10^{18} {\rm m^2}$, 
$\tau$ takes range between $\tau=10^{-5}-10^3$, depending on the size of particles between $r_{\rm p}=10^{-3}-10^5$ m. 
$\epsilon$ is the coefficient of restitution and takes range between $0-1$ depending on the material properties and we use $\epsilon=0.1$ for our calculation.  
$\ln \Lambda$ takes range between $1-10$, respectively. 
The coefficients $c_1$ and $c_2$ are of order unity and depend on $\tau$ \citep{GT78} and/or spin state of particles \cite{Mor06}. 
The dynamical evolution of the debris can be divided into three stages that we will discuss in detail in the following subsections. 
At each stage, we compare timescales of accretion, damping and spreading.

\subsection{Collisional damping of the initial hot debris}
\label{1st_stage}
\subsubsection{Collisional damping timescale}
\begin{figure}[ht!]
\figurenum{5}
	\epsscale{0.65}	
	\plotone{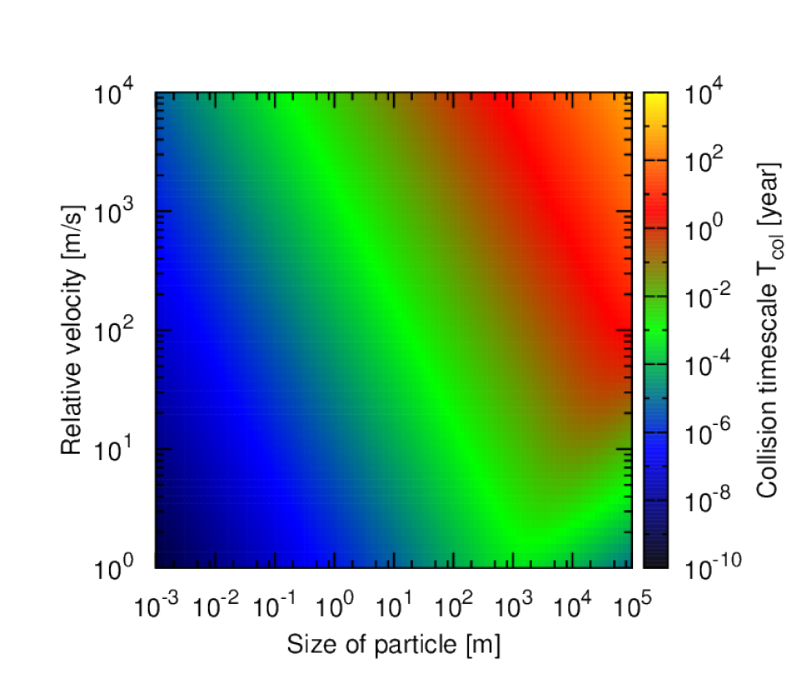}
 \vspace{0mm}
\caption{Collision timescale as a function of particle size and velocity dispersion (Eq.(\ref{tau_col_eq})).}
\label{tau_col}
\end{figure}

After the giant impact, the velocity dispersion of particles is much larger than their escape velocity and their shear velocity. 
Initially, the accretion is prohibited. Instead, collisional damping is effective and velocity dispersion gradually decreases. 
In the particle-in-a-box approximation, the collision timescale is written as 
\begin{equation}
 	T_{\rm col} = \frac{1}{n \sigma_{\rm col} v_{\rm rel}}
\label{tau_col_eq}
\end{equation}
where $n$ is the number density of particles, $\sigma_{\rm col}$ is the collisional cross section and $v_{\rm rel} \sim c$ is the relative velocity. The cross section is written as 
\begin{equation}
	\sigma_{\rm col} = \pi r_{\rm p}^2 \left( 1 + v_{\rm esc}^2/v_{\rm rel}^2 \right). 
\end{equation}
Considering the particles are distributed toroidally after the impact, the volume of this toroid can be expressed as $V=(2\pi a)\cdot(\pi a\langle e \rangle a\langle I \rangle)=2\pi^2 a^3 \langle e \rangle \langle I \rangle$, 
assuming radial and vertical widths are $a\langle e \rangle $and $a \langle I \rangle$, respectively, where $\langle e \rangle $and $\langle I \rangle$ are the mean eccentricity and inclination, respectively.  
Thus, the number density is written as 
\begin{equation}
	n = N/V=\frac{N}{2\pi^2a^3 \langle e \rangle \langle I \rangle}
\end{equation}
where we assume that $\langle e \rangle \sim \sqrt{\frac{5}{3}} \langle v_{\rm rel}/v_{\rm K} \rangle $ and $\langle I \rangle \sim \sqrt{\frac{1}{3}} \langle v_{\rm rel}/v_{\rm K} \rangle$, where $v_{\rm K}$ is the Keplerian velocity \citep[see also][]{Jac12}.
Figure \ref{tau_col} shows collision timescale as a function of $r_{\rm p}$ and velocity dispersion $v_{\rm rel} \sim c$. Timescale varies significantly depending on the size of particle and relative velocity. 
We will compare this timescale to viscous spreading timescale in the next subsection.\\

\subsubsection{Spreading timescale without gravitational instability}
\begin{figure}[h!]
\figurenum{6}
	\epsscale{0.65}	
	\plotone{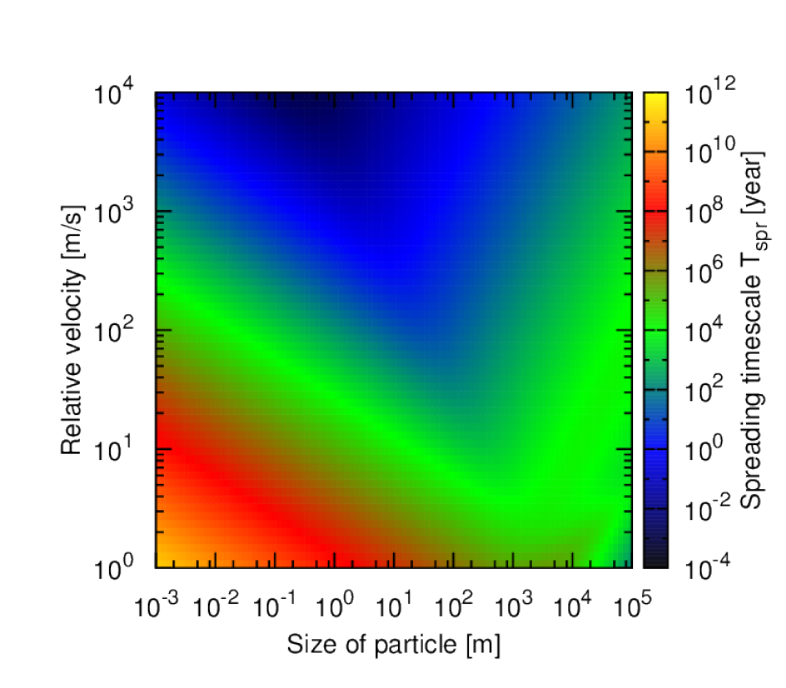}
 \vspace{0mm}
\caption{Spreading timescale when $Q>1$ as a function of particle size and velocity dispersion (Eq.(\ref{tau_spr_Qbig})).}
\label{tau_spr}
\end{figure}

\begin{figure}[h!]
\figurenum{7}
	\epsscale{0.65}	
	\plotone{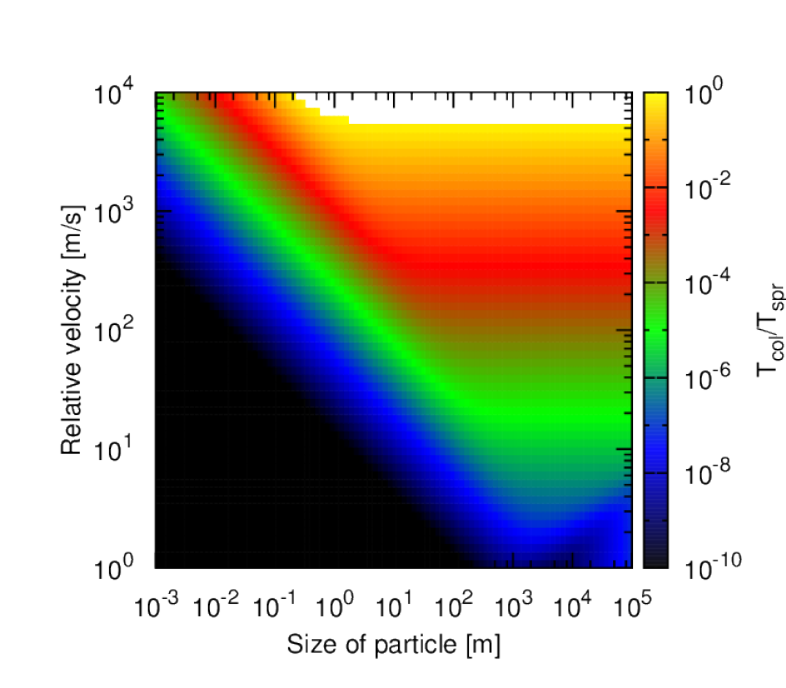}
 \vspace{0mm}
\caption{Ratio of collision timescale to spreading timescale $T_{\rm col}/T_{\rm spr, Q>1}$ when $Q>1$. In the case $(T_{\rm col}/T_{\rm spr, Q>1})>1$, it is plotted with white and 
	in the case $(T_{\rm col}/T_{\rm spr, Q>1})<1\times10^{-10}$, it is plotted with black.}
\label{col_vs_spr}
\end{figure}

As the velocity dispersion decreases, the system may viscously spread.  The timescale of viscous spreading can be written as $T_{\rm spr}=\Delta a^2/\nu$, where $\Delta a$ is the diffusion width and $\nu$ is viscosity, respectively. 
The value of viscosity depends on Toomre's Q parameter \citep{Too64}
\begin{equation}
Q=\frac{c_{\rm r}\kappa}{3.36G\Sigma}
\label{Qpara}
\end{equation} 
where $c_{\rm r}$ is the velocity dispersion in the radial direction and $\kappa$ is the epicyclic frequency, respectively. Initially, $Q$ is much larger than 1 and thus gravitationally stable.
Therefore, the viscosity can be expressed as 
\begin{equation}
	\nu_{\rm Q>1} = \nu_{\rm trans} + \nu_{\rm col}
\label{nu_spr_Qbig}	
\end{equation}
where $\nu_{\rm trans}$ is the translational viscosity \citep{GT78} 
\begin{equation}
	\nu_{\rm trans} = \frac{c^2}{\Omega} \frac{\tau}{1+\tau^2}
\end{equation}
and $\nu_{\rm col}$ is the collisional viscosity \citep{Ara86} 
\begin{equation}
	\nu_{\rm col} = \Omega r_{\rm p}^2 \tau,
\end{equation}
respectively. Then, spreading timescale $T_{\rm spr, Q>1}$ can be written as 
\begin{equation}
	T_{\rm spr, Q>1} = \frac{\Delta a^2}{\nu_{\rm Q>1}}.
\label{tau_spr_Qbig}
\end{equation} 
Figure \ref{tau_spr} shows spreading timescale when $Q>1$ (Eq.(\ref{tau_spr_Qbig})) as a function of velocity dispersion and size of particle. 
Figure \ref{col_vs_spr} shows the ratio of collision timescale to spreading timescale $T_{\rm col}/T_{\rm spr, Q>1}$.  
We find that the collisional damping significantly dominates over the spreading in most of the parameter space considered here ($r_{\rm p}=10^{-3}-10^5$ m and $c=1-10^4$ m s$^{-1}$). 
Thus, the initial hot debris disk is expected to flatten without significant spreading and accretion.

 \newpage
\subsection{Accretion under gravitational instability}
\label{2nd_stage}
\subsubsection{Spreading timescale with gravitational instability}
\begin{figure}[ht!]
\figurenum{8}
	\epsscale{0.5}	
	\plotone{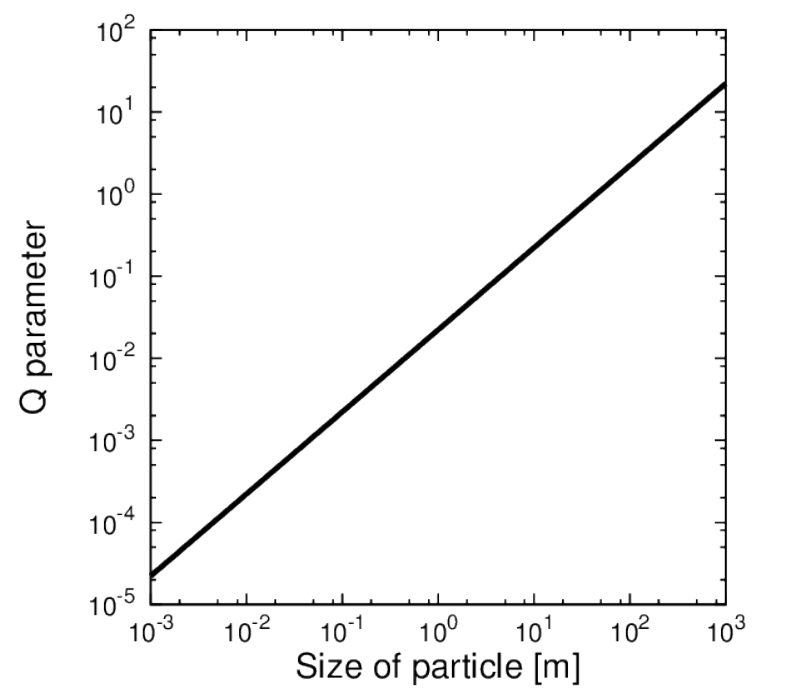}
 \vspace{0mm}
\caption{Toomre's $Q$ parameter as a function of  the size of particle, assuming $c_{\rm r}=v_{\rm esc}$ in Eq. (\ref{Qpara}).}
\label{Qvalue}
\end{figure}

The ratio of the size of Hill sphere to the sum of the particle radii is written as \citep[see also][]{Hyo14}
\begin{equation}
	\tilde{r}_{\rm H} = \frac{R_{\rm H}}{2r_{\rm p}} = 0.82 \left(  \frac{\rho}{900 {\rm kg\hspace{0.1em} m^{-3}}} \right) ^{1/3} \left( \frac{a}{\rm 100,000km} \right)
\end{equation}
where $\rho$ is the density of particle. Using $\rho=1200$ kg m$^{-3}$ and $a=500,000$ km, we get $\tilde{r}_{\rm H} \sim 5$. 
As discussed above, the initial velocity dispersion decreases due to the collisional damping. 
Once the velocity dispersion becomes small enough, gravitational scattering becomes effective and increases the velocity dispersion. 
When $\tilde{r}_{\rm H} > 0.5$, the velocity dispersion at the steady state becomes comparable to the escape velocity of particles \citep{Sal95,Oht99} as
\begin{equation}
	c \sim v_{\rm esc} = \sqrt{\frac{2Gm_{\rm p}}{r_{\rm p}}} = 0.07 {\rm m \hspace{0.1em} s^{-1}} \left( \frac{\rho}{900 {\rm kg \hspace{0.1em} m^{-3}}} \right)^{1/2} \left( \frac{r_{\rm p}}{100 {\rm m}} \right)
\end{equation}
where $m_{\rm p}$ is the particle mass. In this second stage, the $Q$ parameter can become small. Figure \ref{Qvalue} shows the value of Toomre's $Q$ parameter, assuming $c_{\rm r} = v_{\rm esc}$ and $\kappa=\Omega$ with $a=500,000$ km.  
We find that $Q$ becomes smaller than $1$ when particle radius $\lesssim 100$ m (Note that, $r_{\rm p}=100$ m corresponds to $\tau=0.01$ in our work). Therefore, when $r_{\rm p} \lesssim 100$ m, gravitational instability occurs ($Q<1$). 
In this case, the gravitational viscosity dominates over that of collision, and the viscosity can be expressed as \citep{Dai01}
\begin{equation}
	\nu_{\rm Q<1} = 2\nu_{\rm grav} = 52 \tilde{r}_{\rm H}^5 \frac{G^2 \Sigma^2}{\Omega^3}.
\label{nu_spr_Qsmall}
\end{equation}
Thus, spreading timescale can be written as 
\begin{equation}
	T_{\rm spr, Q<1}= \frac{\Delta a^2}{\nu_{\rm Q<1}} \sim 
	1\times10^4 {\rm year} \left( \frac{\Delta a}{3.5\times10^5 {\rm km}} \right)^2 \left( \frac{\Sigma}{2000 {\rm kg \hspace{0.1em}m^{-2}}} \right)^{-2} 
	\left( \frac{\Omega}{2\times10^{-5} {\rm s}} \right)^3 \left( \frac{\tilde{r}_{\rm H}}{5} \right)^{-5}.
\label{tau_spr_Qsmall}
\end{equation}
Using $\tilde{r}_{\rm H}=5$, $\Sigma=M_{\rm tot}/(2\pi a \Delta a_{\rm ini}) \sim 2000$ kg m$^{-3}$, $\Omega \sim 2\times10^{-5}$ s and $\Delta a=3.5\times10^8$ m, we get $T_{\rm spr, Q<1} \sim 1\times10^4$ year. 
Compared to the case of $Q>1$ (see Figure \ref{tau_spr} and Eq.(\ref{tau_spr_Qbig})), the spreading timescale is significantly shorter for this small velocity dispersion (comparable to escape velocity). 
Thus, spreading may occur in this second stage. Next, we will compare this timescale to accretion timescale in the next subsection.

\newpage
\subsubsection{Accretion timescale}
Since the velocity dispersion is now small, particles can accrete and grow. As discussed above, once particle becomes larger than $100$ m sized body, the system becomes gravitationally stable ($Q>1$ and see Figure \ref{Qvalue}). 
Accretion timescale to grow up to the size of $R$ and mass $M$ is written as $T_{\rm grow} = M/\dot{M}$ and the growth rate $\dot{M}$ can be expressed as mass that swept up per unit time as 
\begin{equation}
	\dot{M} = (\Sigma/H) v_{\rm rel} \pi R^2 \left(1 + F_{\rm grav}  \right)
\end{equation}
where $H$ is the scale hight and written as $H=v_{\rm rel}/\Omega$ and $F_{\rm grav}=v_{\rm esc}^2/v_{\rm rel}^2$ is the gravitational focusing factor, respectively. Thus, growth timescale becomes 
\begin{equation}
	T_{\rm grow}  \sim \frac{\rho R}{\Omega \Sigma \left( 1+ F_{\rm grav} \right)} = 
	0.07 {\rm year} \frac{1}{1+F_{\rm grav}} \left( \frac{\rho}{900 {\rm kg \hspace{0.1em}m^{-3}}} \right) 
	\left( \frac{R}{100 {\rm m}} \right) \left( \frac{\Omega}{2\times10^{-5} {\rm s}} \right)^{-1} \left( \frac{\Sigma}{2000 {\rm kg \hspace{0.1em} m^{-2}}} \right)^{-1}
\label{tau_grow}
\end{equation}
Timescale to grow up to $100$ m sized body is independent on the initial size of particle as seen Eq. (\ref{tau_grow}), and considering $\rho=1200$ kg m$^{-3}$, $R=100$ m, $\Sigma=2000$ kg m$^{-2}$, $\Omega=2\times10^{-5}$ s 
and $F_{\rm grav}=1$, we get $T_{\rm grow} \sim 0.05 $ year, which is much shorter than that we obtained for spreading with $Q<1$ (Eq. (\ref{tau_spr_Qsmall})). 
Therefore, accretion takes place quickly without significant spreading even under the gravitational instability and form particles larger than $100$ m. Thus, the system again becomes gravitational stable ($Q>1$).

\subsection{Accretion under gravitational stability}
\label{3rd_stage}
As discussed above, once typical size of particle becomes larger than $100$ m at $a=500,000$  km, the system becomes gravitational stable. Thus, the spreading timescale is regulated by $T_{\rm spr,Q>1}$ (Eq. (\ref{tau_spr_Qbig})) 
which is much longer than the accretion timescale of $T_{\rm grow} \sim$500 years even for 1000 km body with using Eq. (\ref{tau_grow}). \\ 

\cite{Cuk16} assumes that the system is always gravitational instable ($Q<1$) and estimates the spreading timescale by using Eq. (\ref{tau_spr_Qsmall}) as about $2000$ year, which is comparable to the timescale to 
form 1000 km sized object (Eq. (\ref{tau_grow})). Then, they proposed that  the debris may spread all the way inside the Roche limit and form Saturn's rings \citep{Cuk16}. 
However, as we have shown above, the system is rather expected to accrete into several large objects without significant spreading. 
This is also confirmed by $N$-body simulations (Section \ref{N-body}) in the case where we start with large particles ($Q>1$).\\

\section{Conclusion \& Discussion}
\label{Conclusion}

Several scenarios exist for the origin of Saturn's rings. Rings may form during the gas accretion phase ($\sim$ 4.5 Gyrs ago) by tidal disruption of a gas-driven inward-migrating primordial satellite \citep{Can10} or it may have formed during LHB ($\sim$ 3.8 Gyrs ago) by 
tidal disruption of passing large KBOs \citep{Hyo17}. In contrast, rings could be much younger than the Solar system \citep{Cuz98}. Recently, \cite{Cuk16} proposed that Saturn's moon system has experienced a catastrophic impact between 
Rhea-sized objects about 100 Myrs ago around its today's location and that the disk of debris may spread all the way inward to form rings.  They also proposed that current eccentricity of Titan could be induced by the orbital resonance with small 
moons that formed at the edge of the disk and migrate outward due to the interaction with spreading disk.\\

In this paper, using both direct numerical simulations and analytical arguments, we investigated the hypothesis that is proposed in \cite{Cuk16}. 
First, we performed SPH simulations of giant impact between Rhea-sized objects with an impact velocity of $3$ km s$^{-1}$. 
We found that outcome of collision, if catastrophic (for impact angle $45$ degrees), in general form only two large remnants containing about 40\% of the initial total moons' mass. 
These fragments are embedded in a debris disk (Section \ref{SPH}). 
Then, we performed $N$-body simulations using the data obtained from SPH simulations to investigate the longer-term evolution of the debris disk (Section \ref{N-body}). 
$N$-body simulations suggest that the system quickly re-accretes into a single object without significant spreading of the debris.\\

However, in the $N$-body simulations, the effect of fragmentation is not included. After giant impact, the debris particles have large eccentricities and thus successive collisional grinding may occur. In addition, the size of fragments depends on the 
impact angle even though the impact velocity is same (see Fig \ref{largest_rem}). Thus, using analytical arguments, we investigate the fate of the debris in the case they consist of only small particles (Section \ref{analysis}). 
We find that the system follows three different stages of dynamical evolution. Just after the impact, the system is significantly excited.  At this time, Toomre's $Q$ parameter is larger than $1$ and thus the viscosity of the debris is written as 
$\nu_{\rm Q>1} = \nu_{\rm trans} + \nu_{\rm col}$ (Eq. (\ref{nu_spr_Qbig})).  At this first stage, collision damping dominates over viscous spreading. Therefore, the system flattens until the velocity dispersion becomes comparable to 
the particle's escape velocity (Section \ref{1st_stage}). Second, when the velocity dispersion becomes comparable to the escape velocity, the $Q$ parameter can become smaller than $1$ as long as radius of particles is smaller than $100$ m. Under this condition, the 
viscosity is regulated by gravitational interaction as $\nu_{\rm Q<1} = 2\nu_{\rm grav}$ (Eq.(\ref{nu_spr_Qsmall})). Then, we calculated accretion timescale up to $100$ m sized body and we found that the accretion timescale is much shorter than that of 
spreading timescale. Therefore, at this second stage, the accretion dominates over the spreading (Section \ref{2nd_stage}). After particles grow to sizes larger than $100$ m, the system becomes $Q>1$ again. Thus, the viscous spreading is regulated by 
$\nu_{\rm Q>1}$. Comparing the timescale of viscous spreading to accretion timescale to $1000$ km sized body, the accretion timescale is again much shorter than the spreading timescale as long as the velocity dispersion is comparable or smaller than 
the escape velocity of particles. Thus, at this third stage, the accretion further takes place without significant spreading of the system (Section \ref{3rd_stage}).\\

We find that the impact between the two moons is indeed catastrophic as suggested by \cite{Cuk16}. However, we do not find significant spreading, but rather rapid re-acretion of the system. Difference from \cite{Cuk16} comes from the viscosity formula that is used. \cite{Cuk16} assumes that 
the system is always gravitationally instable ($Q<1$) and applied the formula $\nu_{\rm Q<1}$ to estimate the spreading timescale to compare the accretion timescale up to $1000$ km body.  However, as we have shown above, the system is mostly gravitationally stable ($Q>1$) and $\nu_{\rm Q>1}$ should be considered.\\

In conclusion, this study shows that the debris is expected to re-accrete very quickly to form a new-Rhea or/and new-Dione and that spreading is very inefficient after the impact and before complete re-accretion. Therefore, as discussed above, the disk hardly spreads to form Saturn's rings. Thus, the origin of Titan's current eccentricity by disk-driven migration of small moons into orbital resonance with Titan as suggested by \cite{Cuk16} is also less likely to occur.

\acknowledgments
We thank H. Genda for his kindly providing us a SPH code. R.H. thank Shugo Michikoshi for discussion. We also thank L. W. Esposito for useful comments on the manuscript. This work was supported by JSPS Grants-in-Aid for JSPS Fellows (17J01269). 
Part of the numerical simulations were performed using the GRAPE system at the Center for Computational Astrophysics of the National Astronomical Observatory of Japan. 
Also, numerical computations were partly performed on the S-CAPAD platform, IPGP, France.
We acknowledge the financial support of the UnivEarthS Labex programme at Sorbonne Paris Cit{\'e} (ANR-10-LABX-0023 and ANR-11-IDEX-0005-02). This work was also supported by Universit{\'e} Paris Diderot and 
by a Campus Spatial grant. S{\'e}bastien Charnoz thanks the IUF (Institut Universitaire de France) for financial support.
 

\end{document}